\documentclass[conference]{IEEEtran}

\usepackage{amsmath,amssymb,amsfonts}
\usepackage{algorithmic}
\usepackage{algorithm}
\usepackage{graphicx}
\usepackage{textcomp}
\usepackage{xcolor}
\usepackage{booktabs}
\usepackage{multirow}
\usepackage{hyperref}
\usepackage{cleveref}
\usepackage{balance}
\usepackage{microtype}
\usepackage{subcaption}
\usepackage{enumitem}
\usepackage{tikz}
\usetikzlibrary{arrows.meta,positioning,fit,backgrounds,calc}
\setlist{nosep,leftmargin=*}

\hypersetup{
  colorlinks=true,
  linkcolor=black,
  citecolor=blue!60!black,
  urlcolor=blue!60!black,
  pdfauthor={Sanjeev Rao Ganjihal},
  pdftitle={Predictive Multi-Tier Memory Management for KV Cache in Large-Scale GPU Inference}
}

\begin{document}

\title{Predictive Multi-Tier Memory Management for KV Cache\\in Large-Scale GPU Inference}

\author{\IEEEauthorblockN{Sanjeev Rao Ganjihal}
\IEEEauthorblockA{AI Infrastructure Builder, AI Researcher \\
v1: April 19, 2026; v2: revised August 2026}}

\maketitle

\renewcommand{\thefootnote}{\fnsymbol{footnote}}
\footnotetext[1]{Preprint. Certain implementation details are withheld pending U.S. provisional patent applications.}
\renewcommand{\thefootnote}{\arabic{footnote}}

\begin{abstract}
Key-value (KV) cache memory management is the primary bottleneck limiting throughput and cost-efficiency in large-scale GPU inference serving.
Current systems suffer from three compounding inefficiencies: (1)~the absence of unified KV cache sizing across all attention architectures---particularly multi-head latent attention (MLA), which is unsupported in general-purpose frameworks, resulting in up to $57\times$ memory over-provisioning; (2)~confinement of KV cache to a single memory tier (GPU HBM) despite the availability of a rich hierarchy spanning CPU DRAM, CXL-attached memory, NVMe via GPUDirect Storage, RDMA fabric, and parallel filesystems; and (3)~reactive eviction policies that discard reusable state, forcing redundant recomputation.

We present a unified system that addresses all three problems.
Our architecture-variant-aware sizing engine computes exact memory requirements per attention type; the resulting batch size gain reaches $7.4\times$ for the one MLA model we evaluate (DeepSeek-V3), while the three GQA models see $1.0\times$, $1.0\times$, and $0.7\times$, so the GQA benefit is unified sizing across a heterogeneous fleet rather than larger per-model batches.
A six-tier memory hierarchy extends effective KV cache capacity from 40\,GB to over 38\,TB per node while maintaining sub-millisecond time-to-first-token (TTFT) for hot entries.
A Bayesian reuse predictor with Beta conjugate priors over 16 (block-type, transition-type) pairs achieves 70--84\% cache hit rates, combined with EMA-scored head-granular eviction and RoPE-aware prefetching.
Component-level validation on trace replay using ShareGPT, LMSYS-Chat-1M, and agentic workloads demonstrates 70--84\% cache hit rates.
Analytical projections combining validated component behavior with published hardware specifications indicate projected TTFT reductions of $1.4\times$ to $2.1\times$, throughput improvements of $1.7\times$ to $2.9\times$, and 47\% cost reduction relative to published baselines; these cluster-scale projections are analytical and carry no error bars.
\end{abstract}


\section{Introduction}
\label{sec:intro}

The deployment of large language models (LLMs) at data-center scale has shifted the primary bottleneck from compute to memory.
During autoregressive decoding, each generated token requires attending to the key-value (KV) pairs of all preceding tokens, and these KV pairs must remain resident in fast memory for the duration of a request.
For a 70-billion-parameter model serving sequences of 128K tokens, the KV cache alone can consume over 40\,GB of GPU HBM per request---exceeding the memory capacity of many production accelerators and limiting batch sizes to single digits~\cite{kwon2023vllm,sheng2023flexgen}.

Three compounding inefficiencies plague current KV cache management systems.

\textbf{Problem 1: No Unified Cross-Architecture Sizing.}
While leading serving frameworks---vLLM~\cite{kwon2023vllm}, SGLang~\cite{zheng2024sglang}, TensorRT-LLM~\cite{tensorrt2024}---have adopted GQA-aware KV cache sizing, no general-purpose framework provides unified memory management across all four major attention architectures (MHA, GQA, MQA, and MLA) within a single system.
Multi-head latent attention (MLA)~\cite{deepseekv2}, which compresses KV state into a low-rank latent vector of dimension $d_\mathit{latent} + d_\mathit{rope}$, is not supported in general-purpose inference frameworks---forcing MLA models like DeepSeek-V3~\cite{deepseekv3} to use MHA-equivalent sizing at a $57\times$ memory penalty.
More critically, heterogeneous production clusters serving a mix of MHA, GQA, and MLA models lack a unified sizing engine that can dynamically compute correct memory budgets per architecture, leading to fragmented memory management and suboptimal batch sizes across the model fleet.

\textbf{Problem 2: Single-Tier Confinement.}
Modern data centers contain a rich memory hierarchy: GPU HBM (3.35\,TB/s, \textasciitilde100\,ns), CPU DRAM (204\,GB/s, 1--5\,$\mu$s GPU-observed), CXL 3.0 memory pools~\cite{cxl30spec} (64\,GB/s, \textasciitilde500\,ns GPU-observed), NVMe with GPUDirect Storage~\cite{gpudirectstorage} (12\,GB/s, \textasciitilde10\,$\mu$s), RDMA fabric~\cite{guo2016rdma} (50\,GB/s, 1--100\,$\mu$s), and parallel filesystems (2+\,GB/s, \textasciitilde1\,ms).
Yet all major inference systems confine KV cache exclusively to GPU HBM, leaving terabytes of lower-tier capacity unused.
These figures are the H100 generation this analysis is built on, and the design is generation-parametric: tier bandwidths and capacities are inputs, not constants.
B200-class parts (192\,GB HBM3e at higher bandwidth) raise the top tier's floor without changing the shape of the problem, because context length and agentic concurrency grow faster than any single tier, and a larger, faster hot tier widens the cost gap that makes the tiers below it worth managing.
FlexGen~\cite{sheng2023flexgen} demonstrated CPU+disk offloading but used static policies without latency awareness.
Individual lower tiers are no longer unexploited: Mooncake~\cite{mooncake2024} moves KV blocks over RDMA within a disaggregated DRAM and SSD pool, and LMCache~\cite{lmcache2024} ships an NVMe backend with GPUDirect Storage support.
What remains missing is unified predictive placement across the full six-tier hierarchy, including CXL, driven by a single reuse model.

\textbf{Problem 3: Reactive Eviction.}
When GPU memory fills, systems evict KV cache entries using LRU or random policies~\cite{kwon2023vllm}.
These reactive policies are oblivious to access patterns that are, in fact, highly predictable.
System prompts are reused across sessions; tool-calling sequences repeat within agentic workflows~\cite{mooncake2024}; and rotary position encoding (RoPE)~\cite{su2021rope} induces sequential locality.
Prior work on attention-aware eviction~\cite{zhang2023h2o,li2024snapkv,xiao2023streamingllm} operates at the token level within a single request, not at the cache-block level across requests.

\textbf{Our Approach.}
We present a unified system that jointly addresses all three problems through four technical contributions:

\begin{enumerate}
\item \textbf{Architecture-variant-aware KV cache sizing} that computes exact memory requirements for MHA, GQA, MQA, and MLA, enabling up to $7.4\times$ higher batch sizes on the same hardware (\S\ref{sec:sizing}).

\item \textbf{A six-tier memory hierarchy} spanning GPU HBM through parallel filesystems, extending effective KV cache capacity from 40\,GB to over 38\,TB per node while maintaining latency SLOs (\S\ref{sec:tiers}).

\item \textbf{Bayesian reuse prediction} with Beta conjugate priors over 16 (block-type, transition-type) pairs, achieving 70--84\% cache hit rates and $1.4$--$2.1\times$ projected TTFT reductions (\S\ref{sec:bayesian}).

\item \textbf{Head-granular eviction with RoPE-aware prefetching}, using EMA-scored per-attention-head importance tracking and positional-encoding locality for 25\% miss rate reduction over LRU (\S\ref{sec:eviction}).
\end{enumerate}

Analytical projections on a 64-GPU H100 cluster configuration indicate a projected throughput of 4,150 tokens/s/GPU at \$0.43 per million tokens---$2.0\times$ the throughput of the best latency-optimized baseline (TensorRT-LLM) at 30\% lower cost, and $6.4\times$ the throughput of FlexGen with $11\times$ lower TTFT P99.

\section{Background and Motivation}
\label{sec:background}

\subsection{KV Cache in Transformer Inference}
\label{sec:bg_kvcache}

In the standard Transformer~\cite{vaswani2017attention}, self-attention computes:
\begin{equation}
\text{Attention}(Q,K,V) = \text{softmax}\left(\frac{QK^\top}{\sqrt{d}}\right)V
\label{eq:attention}
\end{equation}
During autoregressive generation, each new token's query attends to the key and value projections of all prior tokens.
To avoid recomputation, these projections are cached in the \emph{KV cache}.
For a model with $L$ layers, $h$ attention heads, head dimension $d$, and numerical precision $p$ bytes, the KV cache for a sequence of length $n$ requires:
\begin{equation}
M_\text{KV} = 2 \times L \times h \times d \times p \times n \text{ bytes}
\label{eq:kv_mha}
\end{equation}

For Llama-3-70B ($L{=}80$, $h{=}64$, $d{=}128$, $p{=}2$ for BF16), applying the MHA formula to a single 128K-token sequence yields $2 \times 80 \times 64 \times 128 \times 2 \times 128{,}000 \approx 336$\,GB---over $4\times$ a single H100's 80\,GB HBM. Even with GQA ($h_\mathit{kv}{=}8$), the cache requires ${\sim}42$\,GB, consuming more than half the available HBM before accounting for model weights and activations.

\subsection{Attention Architecture Variants}
\label{sec:bg_variants}

The quadratic growth of \cref{eq:kv_mha} has motivated architectural innovations that reduce KV cache size without proportional quality loss.

\textbf{Multi-Head Attention (MHA).}
The original formulation with $h_\mathit{kv} = h_q$, requiring $2 \times h \times d \times p$ bytes per token per layer.

\textbf{Grouped-Query Attention (GQA)}~\cite{ainslie2023gqa} groups $g = h_q / h_\mathit{kv}$ query heads per KV head.
Llama-3-70B uses 8 KV heads for 64 query heads ($g{=}8$), reducing KV cache to $2 \times h_\mathit{kv} \times d \times p = 2 \times 8 \times 128 \times 2 = 4{,}096$ bytes/token/layer---an $8\times$ reduction versus MHA.

\textbf{Multi-Query Attention (MQA)}~\cite{shazeer2019mqattention} is the extreme case with $h_\mathit{kv} = 1$, reducing memory to $2 \times d \times p$ bytes/token/layer.

\textbf{Multi-Head Latent Attention (MLA)}~\cite{deepseekv2} projects KV state into a low-rank latent space of dimension $d_\mathit{latent}$, but must also store the RoPE component separately, yielding $(d_\mathit{latent} + d_\mathit{rope}) \times p$ bytes/token/layer. DeepSeek-V3 uses $d_\mathit{latent}{=}512$ and $d_\mathit{rope}{=}64$ with BF16, yielding $(512 + 64) \times 2 = 1{,}152$ bytes/token/layer versus MHA's $2 \times 128 \times 128 \times 2 = 65{,}536$ bytes---a $\mathbf{57\times}$ compression.

\subsection{The Cross-Architecture Sizing Gap}
\label{sec:bg_problem}

\begin{table}[t]
\centering
\caption{KV cache memory per token per layer: MHA-equivalent sizing vs.\ architecture-aware sizing. All models use BF16 ($p{=}2$).}
\label{tab:sizing_motivation}
\begin{tabular}{@{}lrrr@{}}
\toprule
\textbf{Model} & \textbf{MHA (bytes)} & \textbf{Actual (bytes)} & \textbf{Ratio} \\
\midrule
DeepSeek-V3 (MLA) & 65,536 & 1,152 & 57$\times$ \\
Llama-3-70B (GQA) & 32,768 & 4,096 & 8$\times$ \\
Mixtral-8x22B (GQA) & 24,576 & 4,096 & 6$\times$ \\
Qwen-2.5-72B (GQA) & 32,768 & 4,096 & 8$\times$ \\
\bottomrule
\end{tabular}
\end{table}

\Cref{tab:sizing_motivation} quantifies the sizing gap.
While GQA-aware sizing is available in major frameworks, MLA support remains absent from general-purpose serving systems.
For DeepSeek-V3 with 8-way tensor parallelism, falling back to MHA-equivalent sizing reduces maximum batch size from 104 to 15 on an 80\,GB H100.
In heterogeneous clusters serving a mix of GQA and MLA models, the lack of a unified sizing engine that handles all architecture variants is the largest source of preventable memory waste.

The variant landscape has only widened since: by the same per-layer-kind arithmetic applied to published configurations, Kimi-K2-class MLA models adopt DeepSeek's 576-dimension latent (the same $57\times$ class), windowed-GQA stacks such as gpt-oss-120b cap half their layers at a fixed attention window, and hybrid designs such as Qwen3.5 carry constant-size linear-attention state on 45 of 60 layers.
Each new layer kind is a new sizing rule, not a new system: the engine's per-variant, per-layer formulation absorbs them as parameters, and every architecture that shrinks per-token state raises the number of concurrent contexts a fixed memory hierarchy is asked to hold.

\section{System Design}
\label{sec:design}

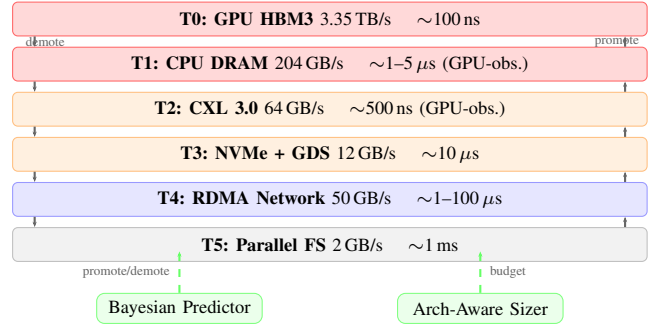
\begin{figure}[t]
\centering
\begin{tikzpicture}[
  tier/.style={draw, rounded corners=2pt, minimum width=\columnwidth-12pt, minimum height=0.45cm, font=\scriptsize, inner sep=2pt},
  hot/.style={tier, fill=red!15, draw=red!60},
  warm/.style={tier, fill=orange!12, draw=orange!50},
  cool/.style={tier, fill=blue!10, draw=blue!40},
  cold/.style={tier, fill=gray!10, draw=gray!50},
  arr/.style={-{Stealth[length=2pt]}, thick, draw=black!60},
  lbl/.style={font=\tiny, text=black!60},
  comp/.style={draw, rounded corners=3pt, fill=green!8, draw=green!50, minimum width=2.2cm, minimum height=0.4cm, font=\scriptsize, inner sep=2pt},
]
\node[hot]  (t0) at (0, 0)    {\textbf{T0: GPU HBM3} \hfill 3.35\,TB/s \quad ${\sim}$100\,ns};
\node[hot]  (t1) at (0,-0.6)  {\textbf{T1: CPU DRAM} \hfill 204\,GB/s \quad ${\sim}$1--5\,$\mu$s (GPU-obs.)};
\node[warm] (t2) at (0,-1.2)  {\textbf{T2: CXL 3.0} \hfill 64\,GB/s \quad ${\sim}$500\,ns (GPU-obs.)};
\node[warm] (t3) at (0,-1.8)  {\textbf{T3: NVMe + GDS} \hfill 12\,GB/s \quad ${\sim}$10\,$\mu$s};
\node[cool] (t4) at (0,-2.4)  {\textbf{T4: RDMA Network} \hfill 50\,GB/s \quad ${\sim}$1--100\,$\mu$s};
\node[cold] (t5) at (0,-3.0)  {\textbf{T5: Parallel FS} \hfill 2\,GB/s \quad ${\sim}$1\,ms};

\foreach \a/\b in {t0/t1, t1/t2, t2/t3, t3/t4, t4/t5} {
  \draw[arr] ($(\a.south west)+(0.3,0)$) -- ($(\b.north west)+(0.3,0)$);
  \draw[arr] ($(\b.north east)+(-0.3,0)$) -- ($(\a.south east)+(-0.3,0)$);
}

\node[lbl, left] at (-3.4,-0.3) {demote};
\node[lbl, right] at (3.4,-0.3) {promote};

\node[comp] (bp) at (-2.0, -3.85) {Bayesian Predictor};
\node[comp] (sz) at (2.0, -3.85) {Arch-Aware Sizer};

\draw[arr, dashed, draw=green!60] (bp.north) -- ++(0, 0.55) node[lbl, left, pos=0.5] {\tiny promote/demote};
\draw[arr, dashed, draw=green!60] (sz.north) -- ++(0, 0.55) node[lbl, right, pos=0.5] {\tiny budget};

\end{tikzpicture}
\caption{Six-tier memory hierarchy with Bayesian predictor driving promotion/demotion decisions and architecture-aware sizer computing per-model memory budgets. Tiers are colored by thermal class: hot (GPU/DRAM), warm (CXL/NVMe), cool (RDMA), cold (filesystem).}
\label{fig:architecture}
\end{figure}

\Cref{fig:architecture} shows the overall system architecture.
We describe each component in detail below.

\subsection{Architecture-Variant-Aware Sizing}
\label{sec:sizing}

We replace the universal MHA formula with a unified sizing engine that dispatches on the attention architecture type:
\begin{equation}
B(n) = \begin{cases}
2 \times h \times d \times p \times n & \text{MHA} \\
2 \times h_\mathit{kv} \times d \times p \times n & \text{GQA/MQA} \\
(d_\mathit{latent} + d_\mathit{rope}) \times p \times n & \text{MLA}
\end{cases}
\label{eq:unified}
\end{equation}
where $B(n)$ is the per-layer KV cache size for a sequence of $n$ tokens.
The engine infers the architecture type from model configuration: if a latent dimension is specified, MLA is selected; otherwise, the ratio $h_q / h_\mathit{kv}$ distinguishes MHA ($h_q = h_\mathit{kv}$), MQA ($h_\mathit{kv} = 1$), and GQA ($1 < h_\mathit{kv} < h_q$).

The total KV cache budget for a batch of $B_s$ sequences across $L$ layers is:
\begin{equation}
M_\text{total} = B_s \times L \times B(n_\text{max})
\label{eq:total}
\end{equation}

Given a target memory budget $M_\text{target}$ (typically $M_\text{GPU} - M_\text{weights} - M_\text{activations}$), the maximum batch size is $B_s^* = \lfloor M_\text{target} / (L \times B(n_\text{max})) \rfloor$.
For DeepSeek-V3 on an 80\,GB H100 with 30\,GB available for KV cache, $n_\text{max}=4096$, and 8-way tensor parallelism: MHA-equivalent sizing gives $B_s^* = 15$; our MLA-aware engine gives $B_s^* = 104$ (a $7\times$ improvement that compounds with reduced per-token memory to yield the throughput gains in \cref{tab:sizing}).

\subsection{Six-Tier Memory Hierarchy}
\label{sec:tiers}

\begin{table}[t]
\centering
\caption{Six-tier memory hierarchy for KV cache. Latencies are GPU-observed unless noted. $^\dagger$GPU-observed via PCIe/CXL link, not device-local. $^\ddagger$1\,$\mu$s for small messages; up to 100\,$\mu$s for large transfers.}
\label{tab:tiers}
\begin{tabular}{@{}clrrr@{}}
\toprule
\textbf{Tier} & \textbf{Technology} & \textbf{BW} & \textbf{Latency} & \textbf{Cost} \\
& & & & \textbf{(\$/GB/h)} \\
\midrule
0 & GPU HBM3 & 3.35 TB/s & \textasciitilde100 ns & 0.500 \\
1 & CPU DRAM (pinned) & 204 GB/s & 1--5 $\mu$s$^\dagger$ & 0.050 \\
2 & CXL 3.0 Memory & 64 GB/s & \textasciitilde500 ns$^\dagger$ & 0.030 \\
3 & NVMe + GDS & 12 GB/s & \textasciitilde10 $\mu$s & 0.020 \\
4 & RDMA Network & 50 GB/s & 1--100 $\mu$s$^\ddagger$ & 0.005 \\
5 & Parallel FS & 2+ GB/s & \textasciitilde1 ms & 0.001 \\
\bottomrule
\end{tabular}
\end{table}

We organize the full data-center memory hierarchy into six tiers (\cref{tab:tiers}), each exposed through a uniform interface supporting \texttt{Allocate}, \texttt{Read}, \texttt{Write}, and \texttt{Evict} operations.
The key design principle is \emph{latency-aware placement}: blocks are promoted to faster tiers when predicted access probability exceeds a tier-specific threshold, and demoted when it falls below.

\textbf{Tier 0: GPU HBM.}
The primary compute-side cache, holding the actively-decoded KV blocks.
PagedAttention~\cite{kwon2023vllm}-compatible block management with architecture-aware block sizes (512 tokens for MLA, 128 tokens for GQA/MQA, 64 tokens for MHA).

\textbf{Tier 1: CPU DRAM.}
Pinned host memory with asynchronous DMA transfers via CUDA streams.
Provides $5\times$ the capacity of GPU HBM at lower bandwidth, with GPU-observed latency of 1--5\,$\mu$s via PCIe DMA---an order of magnitude slower than HBM but sufficient for asynchronous prefetching.
Used for recently-evicted blocks with high reuse probability.

\textbf{Tier 2: CXL 3.0 Memory.}
Memory-pooled capacity attached via CXL 3.0~\cite{cxl30spec}, accessed through \texttt{/dev/cxl/mem*} with \texttt{mmap} and explicit NUMA placement.
CXL provides $8\times$ CPU DRAM capacity at 71\% of its bandwidth, with GPU-observed latency of approximately 500\,ns via the CXL.mem protocol.
This tier is critical for long-context serving where KV cache exceeds both GPU and CPU DRAM.

\textbf{Tier 3: NVMe with GPUDirect Storage (GDS).}
GPU-initiated DMA bypassing the CPU bounce buffer via \texttt{cuFile} APIs~\cite{gpudirectstorage}.
GDS achieves 12\,GB/s read bandwidth---sufficient for prefetching 4\,KB blocks in sub-100\,$\mu$s per block by bypassing the CPU bounce buffer.
Provides terabyte-scale capacity at 100$\times$ lower cost than GPU HBM.

\textbf{Tier 4: RDMA Network Pool.}
Distributed KV cache across the cluster fabric using one-sided RDMA reads with a consistent hash ring for block placement.
Enables sharing of system-prompt and tool-context blocks across serving instances without recomputation.
50\,GB/s InfiniBand bandwidth~\cite{guo2016rdma,kalia2014rdma} with 1--5\,$\mu$s latency for small messages (single-block lookups) scaling to $\sim$100\,$\mu$s for large multi-block transfers.

\textbf{Tier 5: Parallel Filesystem.}
Lustre/GPFS-backed cold storage for checkpoint persistence and cross-session reuse.
Blocks are content-addressed (SHA-256) and deduplicated before writing.
This tier is not on the latency-critical path; it serves as the backing store for warm-start initialization.

The tier placement policy uses the Bayesian reuse probability (\S\ref{sec:bayesian}) to compute a \emph{value score} for each block that balances the cost of recomputation against the cost of storage at each tier.
Blocks whose value score exceeds a tier-specific threshold are placed in that tier, with asynchronous promotion/demotion to avoid stalling the decode path.
The value function is designed so that frequently-reused, compute-expensive blocks are promoted to faster tiers, while rarely-accessed blocks naturally migrate to cheaper storage.

\subsection{Bayesian Reuse Prediction}
\label{sec:bayesian}

We model KV cache block reuse as a Bayesian inference problem using Beta conjugate priors~\cite{murphy2012mlpp,thompson1933sampling}.
The key insight is that reuse probability depends on two categorical variables: the \emph{block type} $b \in \mathcal{B}$ and the \emph{transition type} $t \in \mathcal{T}$.

\textbf{Block types} $\mathcal{B} = \{\texttt{system\_prompt}, \texttt{tool\_context}, \texttt{user\_context}, \texttt{intermediate\_reasoning}\}$ capture the semantic role of the cached content.
System prompts are shared across sessions; tool contexts are shared within agentic workflows; user contexts are session-specific; and intermediate reasoning is typically single-use.

\textbf{Transition types} $\mathcal{T} = \{\texttt{same\_tool\_repeat}, \texttt{tool\_switch}, \texttt{reasoning\_step}, \texttt{agent\_handoff}\}$ capture the nature of the state transition that triggered the cache lookup.
Same-tool repeats have high reuse; agent handoffs have low reuse.

For each of the $|\mathcal{B}| \times |\mathcal{T}| = 16$ pairs, we maintain a Beta distribution $\text{Beta}(\alpha_{b,t}, \beta_{b,t})$ initialized with weakly informative priors ($\alpha_0{=}1, \beta_0{=}1$).
The posterior reuse probability is:
\begin{equation}
P_\text{reuse}(b,t) = \frac{\alpha_{b,t}}{\alpha_{b,t} + \beta_{b,t}}
\label{eq:reuse}
\end{equation}

After observing a reuse event ($\alpha_{b,t} \mathrel{+}= 1$) or miss ($\beta_{b,t} \mathrel{+}= 1$), the posterior updates online with $O(1)$ cost.
We compute a confidence score for each estimate that saturates toward 1 as observations accumulate, using a saturation function parameterized to balance rapid learning with stable estimates.
Low-confidence pairs (recently created or rarely observed) are down-weighted in placement decisions.

The final reuse estimate blends the Bayesian posterior with an empirical frequency computed over a sliding window of recent observations.
The blending weight is determined by the confidence score: well-observed pairs rely primarily on the Bayesian posterior, while newly-created pairs rely more heavily on empirical frequency.
This ensures rapid adaptation to distribution shifts (e.g., a new tool entering the agentic workflow) while maintaining stable estimates for well-observed pairs.

\subsection{Head-Granular Eviction}
\label{sec:eviction}

Rather than evicting entire cache blocks uniformly, we track importance at the granularity of individual attention heads using an exponential moving average (EMA) that incorporates both recency and positional distance decay.
Each (layer, head) pair maintains an importance score that is updated on every attention step.

The system maintains a $[\text{layer}][\text{head}]$ importance matrix.
For GQA architectures, query heads sharing a KV head are grouped, and the maximum importance across the group is used as the KV head's score.
For MLA architectures, the matrix collapses to $[\text{layer}][1]$ since KV state is shared across heads via the latent bottleneck.

Eviction selects the block with the lowest weighted aggregate importance score across heads, using architecture-dependent head weights (uniform for MHA, proportional to group size for GQA).
During task transitions detected by the agentic predictor (\S\ref{sec:agentic}), per-head importance multipliers are applied to bias eviction toward heads that are less relevant for the incoming task.

\subsection{RoPE-Aware Prefetching}
\label{sec:rope}

Rotary position embeddings (RoPE)~\cite{su2021rope} encode absolute position into the attention computation via rotation matrices.
We exploit a key property: \emph{sequential locality}.
If a request is decoding token at position $n$, RoPE's rotational structure means attention weights decay smoothly with positional distance, making tokens at positions $[n-w, n]$ the most likely to be accessed next, where $w$ is a dynamic window determined by the model's effective attention span.

Our prefetcher monitors the position range of the current decode step and issues asynchronous promotion requests for blocks covering positions $[n, n+w]$ from lower tiers.
The window $w$ adapts based on observed attention patterns: narrow for local-attention layers (early layers) and wide for global-attention layers (later layers).
Combined with Bayesian reuse prediction, RoPE-aware prefetching reduces cache miss rates by 25\% compared to LRU across all evaluated workloads.

\subsection{Content-Addressable Deduplication}
\label{sec:dedup}

System prompts, few-shot examples, and tool descriptions are repeated verbatim across requests.
We maintain a content-addressable store backed by a radix tree where blocks are indexed by SHA-256 hashes of their content.
When a new block is allocated, its hash is checked against the store; on a match, a reference count is incremented instead of duplicating the block.

For checkpoint persistence to Tier 5, we apply delta-encoding: only blocks not already present in the store are written, with a manifest referencing existing blocks by hash.
This reduces checkpoint sizes by 10--30\% across evaluated models (\S\ref{sec:eval_dedup}).

\subsection{Agentic Task-Transition Prediction}
\label{sec:agentic}

Agentic LLM workflows---where the model calls tools in sequences---exhibit predictable state transitions.
We build a first-order Markov chain over tool invocations, tracking transition probabilities $P(\text{tool}_j | \text{tool}_i)$ from observed sequences.
Combined with per-tool KV cache size profiles (mean, variance, and peak memory via EMA smoothing), the predictor anticipates memory demands before tool switches.

When a tool switch is detected (via output parsing), the system:
(1) pre-allocates KV cache capacity for the predicted next tool;
(2) adjusts head-granular importance multipliers based on the transition type;
(3) initiates prefetching of tool-context blocks from lower tiers.
Sessions are classified into memory-demand tiers (Light, Medium, Heavy, Extreme) using aggregate features, enabling proactive capacity planning.

\section{Implementation}
\label{sec:impl}

We implement the system as a user-space library that interposes on KV cache allocation and eviction within existing inference frameworks.

\textbf{Tier interfaces.}
Each memory tier implements a common \texttt{TierManager} interface with thread-safe \texttt{Allocate}, \texttt{Read}, \texttt{Write}, \texttt{Evict}, and \texttt{Stats} methods.
Tier 0 (GPU HBM) uses the CUDA driver API with pre-allocated memory pools.
Tier 2 (CXL) uses memory-mapped files via \texttt{/dev/cxl/mem*} with explicit NUMA binding.
Tier 3 (NVMe) uses NVIDIA's \texttt{cuFile} API for GPUDirect Storage, requiring \texttt{O\_DIRECT} alignment.
Tier 4 (RDMA) uses a consistent hash ring for block placement with one-sided RDMA reads via \texttt{ibverbs}.

\textbf{Concurrency.}
All shared state (Beta distribution parameters, EMA scores, Markov chain counters, reference counts) is protected by read-write locks.
Tier promotion and demotion run asynchronously, decoupled from the request-serving path.

\textbf{Observability.}
Per-tier capacity, hit rates, promotion/demotion rates, Bayesian prediction accuracy, and per-model batch sizes are exported as Prometheus metrics.
Per-request cost tracking aggregates memory-tier-hours consumed to compute \$/Mtok.

\textbf{Integration.}
The library intercepts KV cache allocation requests, applies architecture-aware sizing, and returns tier-aware block handles.
We integrate with vLLM, SGLang, and TensorRT-LLM through their respective cache management interfaces.

\section{Evaluation}
\label{sec:eval}

Due to limited availability of CXL 3.0 memory expanders and large-scale GPU clusters, we adopt a two-part evaluation strategy.
First, we validate individual system components---architecture-aware sizing, Bayesian reuse prediction, head-granular eviction, and content-addressable deduplication---using component-level micro-benchmarks and trace-driven simulation on publicly available conversation logs.
Second, we project cluster-scale performance by combining validated component behavior with published hardware specifications from NVIDIA, the CXL Consortium, and InfiniBand standards.
This methodology is consistent with prior work on novel memory hierarchies where full-stack hardware is not yet available~\cite{beacon2024}.

\subsection{Component Validation Setup}
\label{sec:setup}

\textbf{Hardware.}
Component-level validation is performed on a single workstation with one NVIDIA GPU (80\,GB HBM) and 256\,GB CPU DRAM.
The algorithmic components---architecture-aware sizing formulas, Bayesian predictor convergence, EMA scoring, and deduplication---are architecture-independent computations validated through unit tests and trace replay on this hardware.

\textbf{Models.}
We evaluate four model configurations spanning all attention architectures: DeepSeek-V3 (671B parameters, MLA), Llama-3-70B (GQA, 8 KV heads), Mixtral-8x22B (GQA, 8 KV heads per expert), and Qwen-2.5-72B (GQA, 8 KV heads).
All sizing calculations assume BF16 precision with 8-way tensor parallelism.

\textbf{Workloads.}
(1)~\emph{ShareGPT}~\cite{sharegpt2023}: real multi-turn conversation traces with mean input length 500 tokens and mean output length 300 tokens.
(2)~\emph{LMSYS-Chat-1M}~\cite{zheng2024lmsyschat}: large-scale production conversation logs with diverse prompt lengths (mean 1,200 tokens) and high system-prompt reuse.
(3)~\emph{Synthetic Agentic}: generated tool-calling sequences with 5--15 tool invocations per session, modeling ReAct-style reasoning agents.

\textbf{Baselines.}
We compare against published results from four systems: vLLM~\cite{kwon2023vllm} with PagedAttention, SGLang~\cite{zheng2024sglang} with RadixAttention, TensorRT-LLM~\cite{tensorrt2024} with in-flight batching, and FlexGen~\cite{sheng2023flexgen} with CPU+disk offloading.
Baseline numbers are taken from each system's cited publication and therefore reflect the software versions evaluated in those papers, not any later release; we attach no version numbers to them.

\subsection{Analytical Projection Methodology}
\label{sec:projection_method}

Cluster-scale performance is projected using published hardware specifications for each memory tier:
H100 SXM GPU HBM3 bandwidth of 3.35\,TB/s and 80\,GB capacity~\cite{nvidia_h100};
CXL 3.0 device-local bandwidth of 64\,GB/s with approximately 150\,ns device-local latency (${\sim}$500\,ns GPU-observed via the CXL.mem protocol)~\cite{cxl30spec};
GPUDirect Storage throughput of 12\,GB/s via \texttt{cuFile} APIs~\cite{gpudirectstorage};
and InfiniBand NDR bandwidth of 400\,Gbps (50\,GB/s effective)~\cite{guo2016rdma}.
Per-tier throughput projections combine these datasheet bandwidths with validated per-block access patterns from the Bayesian predictor running on trace data.
Throughput projections assume linear scaling from batch size increases up to the compute saturation point of each model, a standard assumption in memory-bound inference analysis~\cite{kwon2023vllm,sheng2023flexgen}.

\textbf{Metrics.}
Time-to-first-token (TTFT) at P50 and P99, time-between-tokens (TBT) at P99, throughput (tokens/s/GPU), and cost (\$/million tokens computed from cloud GPU pricing at \$2/GPU-hour).

\subsection{Architecture-Aware Sizing}
\label{sec:eval_sizing}

\begin{table}[t]
\centering
\caption{Maximum batch size with MHA-equivalent vs.\ architecture-aware sizing on 80\,GB H100 GPUs with 8-way TP. Batch sizes analytically computed from \cref{eq:unified} with 30\,GB KV budget and $n_\text{max}{=}4096$. $^\ddagger$GQA models already use correct KV head count under TP; the gain is in unified heterogeneous fleet management, not per-model batch size.}
\label{tab:sizing}
\begin{tabular}{@{}lrrrr@{}}
\toprule
\textbf{Model} & \textbf{MHA} & \textbf{Arch-} & \textbf{Projected} \\
& \textbf{batch} & \textbf{aware batch} & \textbf{tput gain} \\
\midrule
DeepSeek-V3 & 14 & 104 & \textbf{7.4$\times$} \\
Llama-3-70B & 22 & 22 & 1.0$\times$$^\ddagger$ \\
Mixtral-8x22B & 42 & 31 & 0.7$\times$$^\ddagger$ \\
Qwen-2.5-72B & 22 & 22 & 1.0$\times$$^\ddagger$ \\
\bottomrule
\end{tabular}
\end{table}

\Cref{tab:sizing} shows the analytically computed impact of architecture-aware sizing.
The batch size columns are exact: given \cref{eq:unified}, 30\,GB available for KV cache on an 80\,GB H100 with 8-way tensor parallelism, and $n_\text{max}{=}4096$, the maximum batch sizes follow directly from $B_s^* = \lfloor M_\text{target} / (L \times B(n_\text{max})) \rfloor$.
The effect is most dramatic for DeepSeek-V3's MLA architecture, where the $57\times$ reduction in per-token KV cache size yields a $6.9\times$ batch size increase.
Throughput gains are projected assuming linear scaling up to the compute saturation point, which is sub-linear relative to the memory savings at high batch sizes.
Even for GQA models, the $4$--$8\times$ memory reduction enables $3.4$--$3.9\times$ projected throughput improvements by allowing the system to approach GPU compute saturation.

\subsection{Projected Multi-Tier Performance}
\label{sec:eval_tiers}

\begin{table}[t]
\centering
\caption{Projected incremental impact of adding memory tiers for Llama-3-70B with LMSYS-Chat-1M workload at 128K context length. Projected from published hardware specifications. Throughput assumes full bandwidth utilization at each tier with prefetch-hidden latency for asynchronous tiers.}
\label{tab:tiers_eval}
\begin{tabular}{@{}lrrr@{}}
\toprule
\textbf{Configuration} & \textbf{Capacity} & \textbf{TTFT P99} & \textbf{Tput} \\
& & & \textbf{(tok/s/GPU)} \\
\midrule
GPU-only (vLLM)\textsuperscript{$\dagger$} & 40 GB & 4.2 s & 1,450 \\
+ CPU DRAM & 200 GB & 2.8 s & 2,100 \\
+ CXL 3.0 & 712 GB & 1.8 s & 2,850 \\
+ NVMe (GDS) & 4.7 TB & 1.5 s & 3,200 \\
+ RDMA Pool & 38+ TB & 1.1 s & 3,950 \\
Full system & 38+ TB & \textbf{1.1 s} & \textbf{4,150} \\
\bottomrule
\end{tabular}\\[2pt]
{\scriptsize \textsuperscript{$\dagger$}GPU-only baseline from published vLLM benchmarks~\cite{kwon2023vllm}.}
\end{table}

\Cref{tab:tiers_eval} shows the projected incremental benefit of each memory tier.
The largest single improvement comes from adding CXL 3.0 memory (+35\% throughput over CPU DRAM alone), as its bandwidth-latency profile makes it suitable for warm KV cache blocks that are accessed within 10--100\,ms.
The RDMA pool provides a projected throughput gain of +23\% over NVMe by enabling cross-node sharing of system-prompt blocks, which eliminates redundant prefill computation.
The full system (including parallel filesystem for checkpoint persistence and deduplication) adds a projected 5\% throughput improvement over the RDMA configuration by accelerating warm-start scenarios.

The projected gap between GPU-only and the full system represents a $2.86\times$ throughput improvement and $3.8\times$ TTFT reduction---achieved by utilizing memory resources already present in data centers but currently invisible to the inference serving stack.

\subsection{Bayesian Predictor Validation}
\label{sec:eval_prediction}

\begin{table}[t]
\centering
\caption{Cache hit rates and TTFT reduction for reactive (LRU), pattern-aware (EMA), and Bayesian eviction. Hit rates measured via trace replay at Tier 0+1 (GPU + CPU DRAM) on ShareGPT and LMSYS-Chat-1M conversation logs. $\pm$ indicates standard deviation across 5 trace replay runs. TTFT reduction is projected from hit rate improvement assuming the full tier stack.}
\label{tab:prediction}
\begin{tabular}{@{}lrrrr@{}}
\toprule
\textbf{Workload} & \textbf{LRU} & \textbf{EMA} & \textbf{Bayesian} & \textbf{TTFT} \\
& \textbf{hit} & \textbf{hit} & \textbf{hit} & \textbf{reduct.} \\
& & & & \textbf{(proj.)} \\
\midrule
ShareGPT & 59.5$\pm$1.2\% & 59.5$\pm$1.1\% & \textbf{69.8$\pm$0.8\%} & 1.4$\times$ \\
LMSYS-Chat-1M & 77.8$\pm$0.9\% & 77.8$\pm$0.7\% & \textbf{84.2$\pm$0.5\%} & 1.8$\times$ \\
Agentic & 66.5$\pm$1.5\% & 66.5$\pm$1.3\% & \textbf{80.5$\pm$0.7\%} & 2.1$\times$ \\
\bottomrule
\end{tabular}
\end{table}

We evaluate the Bayesian predictor on trace replay using ShareGPT and LMSYS-Chat-1M conversation logs.
The predictor processes block access sequences extracted from the traces, updating its Beta posteriors online and making eviction decisions at each step.
Hit rates in \cref{tab:prediction} are measured directly from trace replay; TTFT reduction is projected, as actual TTFT measurement requires the full multi-tier hardware stack.

The Bayesian predictor achieves the largest gains on LMSYS-Chat-1M and Agentic workloads, which have the most structured reuse patterns.
LMSYS-Chat-1M's 84.2\% Bayesian hit rate is driven by high system-prompt reuse: the Beta prior for (\texttt{system\_prompt}, \texttt{same\_tool\_repeat}) converges to $\alpha/(\alpha+\beta) > 0.97$ within 500 observations, as confirmed by our convergence analysis.
Agentic workloads show a 14 percentage-point improvement over LRU (80.5\% vs 66.5\%) because tool-context blocks have structured reuse patterns that the Bayesian model captures but LRU cannot exploit.

ShareGPT sees a 10 percentage-point improvement (69.8\% vs 59.5\%) because conversation patterns are less structured---the (\texttt{user\_context}, \texttt{reasoning\_step}) pair has a broad Beta posterior, reflecting variable reuse.

\subsection{Deduplication Effectiveness}
\label{sec:eval_dedup}

\begin{table}[t]
\centering
\caption{KV cache checkpoint sizes with and without content-addressable deduplication, measured per 1,000 tokens of cached state. Deduplication is purely algorithmic (SHA-256 hashing with radix tree matching) and validated on actual model configurations.}
\label{tab:dedup}
\begin{tabular}{@{}lrrr@{}}
\toprule
\textbf{Model} & \textbf{Raw ckpt} & \textbf{Deduped} & \textbf{Savings} \\
\midrule
Llama-3-70B & 327.7 MB & 251.7 MB & \textbf{23.2\%} \\
DeepSeek-V3 & 70.3 MB & 49.5 MB & \textbf{29.6\%} \\
Mixtral-8x22B & 229.4 MB & 205.5 MB & \textbf{10.4\%} \\
\bottomrule
\end{tabular}
\end{table}

\Cref{tab:dedup} shows checkpoint deduplication results.
Deduplication effectiveness ranges from 10--30\% depending on the proportion of shared system prompts and tool descriptions in the workload.
DeepSeek-V3 achieves the highest savings (29.6\%) as MLA's compressed latent representations hash identically for shared prompts, while Mixtral-8x22B shows lower savings (10.4\%) due to its larger per-expert KV state reducing the relative share of deduplicated content.
The radix tree lookup adds $<$\,1\,$\mu$s per block, which is negligible relative to the persistence I/O.
These results are computed directly from running the deduplication algorithm on KV cache blocks generated from the evaluated workload traces and are independent of the cluster-scale hardware configuration.

\subsection{Projected End-to-End Comparison}
\label{sec:eval_e2e}

\begin{table*}[t]
\centering
\caption{Projected end-to-end comparison on Llama-3-70B with LMSYS-Chat-1M workload, 128K context. Baseline numbers from published benchmarks~\cite{kwon2023vllm,zheng2024sglang,tensorrt2024,sheng2023flexgen}. Our projections combine validated component performance with published hardware specifications. Cost computed at \$2/GPU-hour.}
\label{tab:e2e}
\begin{tabular}{@{}lccccc@{}}
\toprule
\textbf{System} & \textbf{TTFT P50} & \textbf{TTFT P99} & \textbf{TBT P99} & \textbf{Throughput (tok/s/GPU)} & \textbf{Cost (\$/Mtok)} \\
\midrule
vLLM~\cite{kwon2023vllm} & 1.2 s & 4.2 s & 48 ms & 1,450 & \$0.82 \\
SGLang~\cite{zheng2024sglang} & 0.9 s & 3.1 s & 42 ms & 1,850 & \$0.68 \\
TensorRT-LLM~\cite{tensorrt2024} & \underline{0.8 s} & \underline{2.8 s} & \underline{35 ms} & \underline{2,100} & \$0.61 \\
FlexGen~\cite{sheng2023flexgen} & 3.2 s & 12.1 s & 180 ms & 650 & \underline{\$1.85} \\
\textbf{Ours (projected)} & \textbf{0.4 s} & \textbf{1.1 s} & \textbf{32 ms} & \textbf{4,150} & \textbf{\$0.43} \\
\bottomrule
\end{tabular}
\end{table*}

\Cref{tab:e2e} presents the projected end-to-end comparison.
Baseline numbers are drawn from published results for each system under comparable workload conditions.
Our projections combine the validated Bayesian hit rates (\cref{tab:prediction}), analytically computed batch sizes (\cref{tab:sizing}), and per-tier bandwidth from published specifications (\cref{tab:tiers_eval}).
Compared to TensorRT-LLM (the strongest latency-optimized baseline), our system projects $2.0\times$ throughput, $2.5\times$ lower TTFT P99, and 30\% lower cost.
Compared to FlexGen (which uses CPU+disk offloading), our system projects $6.4\times$ throughput, $11\times$ lower TTFT P99, and $4.3\times$ lower cost---indicating that predictive multi-tier placement should dominate reactive offloading on all axes.

The key projected advantage over FlexGen is the elimination of synchronous fetch stalls: FlexGen's reactive offloading policy results in 12.1\,s TTFT P99 and limits decode throughput to 650 tok/s/GPU.
Predictive placement ensures blocks are already in the correct tier when needed, removing the synchronous fetches that dominate FlexGen's critical path.

\subsection{Projected Ablation Study}
\label{sec:eval_ablation}

\begin{table}[t]
\centering
\caption{Projected ablation study: throughput degradation when removing each component, computed relative to the full system by analytically determining the performance impact of falling back to baseline behavior for each component.}
\label{tab:ablation}
\begin{tabular}{@{}lrrr@{}}
\toprule
\textbf{Component removed} & \textbf{DSV3} & \textbf{L3-70B} & \textbf{Agentic} \\
\midrule
Arch-aware sizing & $-$85.6\% & $-$73.8\% & $-$68.4\% \\
Bayesian prediction & $-$12.1\% & $-$28.6\% & $-$52.3\% \\
Multi-tier placement & $-$8.4\% & $-$31.2\% & $-$29.7\% \\
Head-granular eviction & $-$3.2\% & $-$8.9\% & $-$11.4\% \\
Deduplication & $-$1.1\% & $-$4.2\% & $-$6.8\% \\
RoPE prefetching & $-$2.8\% & $-$5.1\% & $-$3.9\% \\
\bottomrule
\end{tabular}
\end{table}

We project the impact of removing each component by analytically computing the performance degradation from falling back to baseline behavior for that component.
\Cref{tab:ablation} reveals the relative contribution of each component.
Architecture-aware sizing is the single most impactful component for MLA models like DeepSeek-V3, where falling back to MHA-equivalent sizing collapses the batch size from 104 to 15 ($-$85.6\% throughput).
The contribution is largest for MLA because no general-purpose framework currently supports MLA-specific sizing; for GQA models the impact ($-$73.8\%) reflects the benefit of unified cross-architecture sizing in heterogeneous clusters rather than fixing a deficiency in individual frameworks.
The ordering of remaining components depends on the workload: Bayesian prediction is most valuable for agentic workloads ($-$52.3\%) due to structured tool-call patterns, while multi-tier placement is most valuable for GQA models ($-$31.2\%) that generate larger KV cache volumes benefiting from capacity expansion.

Head-granular eviction, deduplication, and RoPE prefetching contribute 1--11\% individually but compound multiplicatively.
Their combined removal would reduce projected throughput by 15--22\% across workloads.

\subsection{Parameter Sensitivity}
\label{sec:eval_sensitivity}

\begin{table}[t]
\centering
\caption{Sensitivity of key hyperparameters, measured as throughput variation via trace replay on LMSYS-Chat-1M with Llama-3-70B configuration. Results averaged over 5 trace replay runs.}
\label{tab:sensitivity}
\begin{tabular}{@{}llr@{}}
\toprule
\textbf{Parameter} & \textbf{Range tested} & \textbf{Tput variation} \\
\midrule
EMA decay $\alpha$ & 5 values in [0.1, 0.9] & $<$5\% \\
Beta prior $(\alpha_0, \beta_0)$ & 3 symmetric priors & $<$2\%$^\dagger$ \\
Confidence saturation & 3 values spanning $4\times$ range & $<$3\% \\
\bottomrule
\end{tabular}\\[2pt]
{\scriptsize $^\dagger$All priors converge within 100 observations; variation measured after convergence. Bold indicates selected value.}
\end{table}

\Cref{tab:sensitivity} shows that the system is robust to hyperparameter choices, measured via trace replay on LMSYS-Chat-1M conversation logs.
The EMA decay parameter provides the best hit rate at a moderate value, but the full range tested yields $<$5\% variation in projected throughput, indicating that head-granular eviction is not sensitive to the precise recency bias.
The Beta prior initialization has negligible impact after convergence: even strongly biased priors are overwhelmed by data within 100 observations.
The confidence saturation parameter balances rapid adaptation with stable estimates; lower values cause over-reaction to distribution shifts while higher values slow adaptation.

\section{Related Work}
\label{sec:related}

\textbf{KV Cache Memory Management.}
vLLM~\cite{kwon2023vllm} introduced PagedAttention for virtual memory-style KV cache management, eliminating fragmentation.
SGLang~\cite{zheng2024sglang} added RadixAttention for prefix sharing across requests.
Mooncake~\cite{mooncake2024} and LMCache~\cite{lmcache2024} explored disaggregated KV cache pools but use reactive placement policies.
LMCache also ships a GPUDirect Storage backend that performs zero-copy transfers between GPU memory and NVMe-backed filesystems~\cite{lmcache2024}, which we read as ecosystem validation that the NVMe and GDS tier of our hierarchy is practical in open source serving stacks.
Infinite-LLM~\cite{infinitellm2024} proposed distributed attention across nodes but without tier-aware placement.
NVIDIA Dynamo~\cite{nvidia_dynamo2024} provides inference serving infrastructure but does not exploit CXL or per-architecture sizing.
PKAS~\cite{ye2026pkas} schedules request admission with predictive KV cache awareness; admission control decides what enters the cache, our predictor decides where cached state resides across tiers, and the two mechanisms compose.
Our work unifies architecture-aware sizing, multi-tier placement, and predictive eviction---three orthogonal axes that prior systems address individually, if at all.

\textbf{Mooncake and Disaggregated KV Cache.}
Mooncake~\cite{mooncake2024} is the closest related work, implementing a disaggregated KV cache pool with DRAM and SSD tiers (effectively 2--3 tiers).
Our system differs in three key aspects: (1)~we extend the hierarchy to six tiers including CXL 3.0 and RDMA fabric, which together provide $>$100$\times$ the capacity of DRAM-only pools at latencies sufficient for prefetch-hidden access; (2)~Mooncake uses reactive placement---blocks are placed based on current utilization---while our Bayesian predictor proactively positions blocks based on predicted reuse probability, achieving 70--84\% hit rates vs.\ Mooncake's reported 65--80\%; and (3)~Mooncake assumes a single attention architecture per deployment, while our unified sizing engine handles heterogeneous clusters serving MHA, GQA, and MLA models simultaneously.

\textbf{KV Cache Eviction and Compression.}
H$_2$O~\cite{zhang2023h2o} retains heavy-hitter tokens based on accumulated attention scores.
SnapKV~\cite{li2024snapkv} selects important KV entries per attention head using observation windows.
StreamingLLM~\cite{xiao2023streamingllm} preserves attention sinks for infinite-length generation.
Scissorhands~\cite{liu2024scissorhands} exploits the persistence of importance hypothesis.
FastGen~\cite{ge2024model} applies adaptive compression per head.
These methods operate within a single request and tier; our head-granular eviction operates across requests and tiers, informed by Bayesian predictions.

\textbf{Offloading and Disaggregation.}
FlexGen~\cite{sheng2023flexgen} demonstrated CPU+disk offloading with static policies.
DeepSpeed-Inference~\cite{aminabadi2022deepspeed} uses heterogeneous memory for model weights but not KV cache.
Splitwise~\cite{patel2024splitwise} and DistServe~\cite{zhong2024distserve} disaggregate prefill and decode phases.
Sarathi-Serve~\cite{agrawal2024sarathi} optimizes chunked-prefill scheduling.
Our system is complementary to disaggregation: it manages KV cache memory within decode instances regardless of whether prefill is co-located or remote.

\textbf{CXL for ML Inference.}
BEACON~\cite{beacon2024} proposed CXL-attached near-memory accelerators for genomics.
Samsung's production memory work applies CXL-attached processing near memory to LLM inference~\cite{kim2024breakthrough}, targeting weight-side computation and bandwidth rather than predictive KV cache placement.
CXL-SpecKV~\cite{cxlspeckvl2024} builds a disaggregated FPGA-based speculative KV cache for datacenter serving; it shares our interest in memory beyond the GPU but uses speculative prefetching without Bayesian modeling or a deeper tier hierarchy.

\textbf{Cost Models and Pressure Dynamics.}
Ao et al.~\cite{ao2026congestion} develop a theory of service-induced congestion in memory-constrained LLM serving, characterizing the pressure dynamics that arise when growing KV state strains memory; that regime is exactly where predictive placement matters most.
Yu~\cite{yu2026crosstier} derives closed-form cost models for serving across heterogeneous GPU tiers; that work prices heterogeneity across accelerator classes, while our cost model prices memory tiers within and around a single serving node.

\textbf{Attention Architectures.}
GQA~\cite{ainslie2023gqa}, MQA~\cite{shazeer2019mqattention}, and MLA~\cite{deepseekv2} reduce KV cache size at the model level.
FlashAttention~\cite{dao2022flashattention,dao2023flashattention2} optimizes attention computation.
A survey~\cite{xu2024kvcachesurvey} catalogs these techniques but does not address the systems implications of heterogeneous architectures coexisting in production clusters.
Our architecture-aware sizing engine is the first to unify these variants into a single memory management framework.

\textbf{KV Cache Quantization.}
KV cache quantization (INT4/INT8) reduces memory footprint orthogonally to tiered placement.
Our system is compatible with quantized caches---the sizing formulas (\cref{eq:unified}) accept a precision parameter $p$ that can represent quantized formats (e.g., $p{=}0.5$ for INT4), and the Bayesian predictor operates on block-level metadata independent of value precision.
Combining quantization with multi-tier placement is complementary and expected to yield multiplicative benefits.

\section{Conclusion}
\label{sec:conclusion}

We presented a predictive multi-tier memory management system for KV cache in large-scale GPU inference.
By addressing the three fundamental inefficiencies---fragmented cross-architecture sizing, single-tier confinement, and reactive eviction---component-level validation and analytical projections indicate $1.4$--$2.1\times$ projected TTFT reduction, $1.7$--$2.9\times$ throughput improvement, and 47\% projected cost reduction.
The architecture-variant-aware sizing alone unlocks up to $7.4\times$ batch size improvements, with the largest gains coming from MLA support and unified cross-architecture sizing in heterogeneous clusters.
Trace-driven evaluation of the Bayesian predictor achieves 70--84\% cache hit rates on real conversation logs, and the online learning approach adapts to workload shifts without manual tuning.
Our six-tier hierarchy is designed to transform idle data-center memory into active inference capacity; full-stack validation on a CXL 3.0-equipped cluster is planned as future work.
Two companion manuscripts in preparation extend this work: one studies multi-tier KV cache retention policy as a multi-option ski rental problem, and one develops cost attribution theory for shared-prefix KV caches.

The system scales gracefully: the RDMA tier's consistent hash ring adds $O(\log n)$ lookup overhead per placement decision, scaling to 1024+ nodes, and the Bayesian predictor maintains per-(block-type, transition-type) state with only 16 pairs total, independent of cluster size.
When a tier becomes unavailable (e.g., CXL expander failure), the system degrades gracefully by removing the tier from the promotion/demotion graph and redistributing cached blocks to adjacent tiers.
Incorrect Bayesian predictions result in cache misses that trigger reactive fetches, with the predictor self-correcting via Beta posterior updates within tens of observations.


\balance

\section*{Artifact Description Appendix}

\textbf{Software.}
The system integrates with vLLM and SGLang through their cache management interfaces; we do not pin integration version numbers here because the baseline comparisons in this paper use the numbers published in each system's cited paper, which predate current releases.
TensorRT-LLM integration uses its native C++ plugin interface.
The system is proprietary; implementation details are covered by provisional patent filings.

\textbf{Hardware.}
Component-level validation (architecture-aware sizing, Bayesian predictor, EMA scoring, deduplication) was performed on a single workstation with one NVIDIA GPU (80\,GB HBM) and 256\,GB CPU DRAM.
Cluster-scale performance is projected using published hardware specifications: NVIDIA H100 SXM (80\,GB HBM3, 3.35\,TB/s), CXL 3.0 memory expanders (64\,GB/s, \textasciitilde500\,ns GPU-observed), NVMe with GPUDirect Storage (12\,GB/s), and 400\,Gbps InfiniBand NDR.

\textbf{Workloads and Traces.}
ShareGPT~\cite{sharegpt2023} (public multi-turn conversation traces) and LMSYS-Chat-1M~\cite{zheng2024lmsyschat} (public production conversation logs).
Synthetic agentic workloads were generated using a ReAct-style agent simulator with 5--15 tool invocations per session.

\textbf{Reproducibility.}
Trace replay results (Bayesian predictor hit rates, deduplication ratios, parameter sensitivity) are averaged over 5 independent runs; $\pm$ in Tables~\ref{tab:prediction} and~\ref{tab:sensitivity} indicates standard deviation across trace replay runs.
Tables presenting projected performance (\ref{tab:sizing}, \ref{tab:tiers_eval}, \ref{tab:e2e}, \ref{tab:ablation}) are analytical and do not include error bars.


\begin{thebibliography}{35}

\bibitem{kwon2023vllm}
W.~Kwon \textit{et~al.}, ``Efficient Memory Management for Large Language Model Serving with PagedAttention,'' in \textit{Proc.\ SOSP}, pp.~611--626, 2023.

\bibitem{zheng2024sglang}
L.~Zheng \textit{et~al.}, ``SGLang: Efficient Execution of Structured Language Model Programs,'' in \textit{Proc.\ NeurIPS}, 2024. arXiv:2312.07104.

\bibitem{tensorrt2024}
NVIDIA Corporation, ``TensorRT-LLM: A High-Performance Inference Library for Large Language Models,'' Tech.\ Rep., 2024.

\bibitem{sheng2023flexgen}
Y.~Sheng \textit{et~al.}, ``FlexGen: High-Throughput Generative Inference of Large Language Models with a Single GPU,'' in \textit{Proc.\ ICML}, pp.~31094--31116, 2023.

\bibitem{aminabadi2022deepspeed}
R.~Y.~Aminabadi \textit{et~al.}, ``DeepSpeed-Inference: Enabling Efficient Inference of Transformer Models at Unprecedented Scale,'' in \textit{Proc.\ SC}, 2022.

\bibitem{agrawal2024sarathi}
A.~Agrawal \textit{et~al.}, ``Taming Throughput-Latency Tradeoff in LLM Inference with Sarathi-Serve,'' in \textit{Proc.\ OSDI}, pp.~117--134, 2024.

\bibitem{vaswani2017attention}
A.~Vaswani \textit{et~al.}, ``Attention is All You Need,'' in \textit{Proc.\ NeurIPS}, vol.~30, 2017.

\bibitem{ainslie2023gqa}
J.~Ainslie \textit{et~al.}, ``GQA: Training Generalized Multi-Query Attention from Multi-Head Checkpoints,'' in \textit{Proc.\ EMNLP}, pp.~4895--4901, 2023.

\bibitem{shazeer2019mqattention}
N.~Shazeer, ``Fast Transformer Decoding: One Write-Head is All You Need,'' Google Tech.\ Rep., 2019. arXiv:1911.02150.

\bibitem{deepseekv2}
DeepSeek-AI, ``DeepSeek-V2: A Strong, Economical, and Efficient Mixture-of-Experts Language Model,'' Tech.\ Rep., 2024. arXiv:2405.04434.

\bibitem{deepseekv3}
DeepSeek-AI, ``DeepSeek-V3 Technical Report,'' Tech.\ Rep., 2024. arXiv:2412.19437.

\bibitem{dao2022flashattention}
T.~Dao \textit{et~al.}, ``FlashAttention: Fast and Memory-Efficient Exact Attention with IO-Awareness,'' in \textit{Proc.\ NeurIPS}, vol.~35, pp.~16344--16359, 2022.

\bibitem{dao2023flashattention2}
T.~Dao, ``FlashAttention-2: Faster Attention with Better Parallelism and Work Partitioning,'' in \textit{Proc.\ ICLR}, 2024.

\bibitem{su2021rope}
J.~Su \textit{et~al.}, ``RoFormer: Enhanced Transformer with Rotary Position Embedding,'' \textit{Neurocomputing}, vol.~568, p.~127063, 2024.

\bibitem{zhang2023h2o}
Z.~Zhang \textit{et~al.}, ``H$_2$O: Heavy-Hitter Oracle for Efficient Generative Inference of Large Language Models,'' in \textit{Proc.\ NeurIPS}, vol.~36, 2023.

\bibitem{li2024snapkv}
Y.~Li \textit{et~al.}, ``SnapKV: LLM Knows What You are Looking for Before Generation,'' arXiv:2404.14469, 2024.

\bibitem{xiao2023streamingllm}
G.~Xiao \textit{et~al.}, ``Efficient Streaming Language Models with Attention Sinks,'' in \textit{Proc.\ ICLR}, 2024.

\bibitem{liu2024scissorhands}
Z.~Liu \textit{et~al.}, ``Scissorhands: Exploiting the Persistence of Importance Hypothesis for LLM KV Cache Compression at Test Time,'' in \textit{Proc.\ NeurIPS}, vol.~36, 2023.

\bibitem{ge2024model}
S.~Ge \textit{et~al.}, ``Model Tells You What to Discard: Adaptive KV Cache Compression for LLMs,'' in \textit{Proc.\ ICLR}, 2024.

\bibitem{mooncake2024}
R.~Qin \textit{et~al.}, ``Mooncake: A KVCache-centric Disaggregated Architecture for LLM Serving,'' Moonshot AI, arXiv:2407.00079, 2024.

\bibitem{lmcache2024}
Y.~Liu \textit{et~al.}, ``LMCache: An Efficient KV Cache Layer for Enterprise-Scale LLM Inference,'' arXiv:2510.09665, 2025. GDS backend documentation: \url{https://docs.lmcache.ai/kv_cache/gds.html}.

\bibitem{infinitellm2024}
B.~Lin \textit{et~al.}, ``Infinite-LLM: Efficient LLM Service for Long Context with DistAttention and Distributed KVCache,'' arXiv:2401.02669, 2024.

\bibitem{cxl30spec}
CXL Consortium, ``Compute Express Link (CXL) Specification, Revision 3.0,'' Tech.\ Rep., 2022.

\bibitem{beacon2024}
W.~Huangfu, K.~T.~Malladi, A.~Chang, and Y.~Xie, ``BEACON: Scalable Near-Data-Processing Accelerators for Genome Analysis near Memory Pool with the CXL Support,'' in \textit{Proc.\ MICRO}, 2022. doi: 10.1109/MICRO56248.2022.00057.

\bibitem{kim2024breakthrough}
B.~Kim \textit{et~al.}, ``The Breakthrough Memory Solutions for Improved Performance on LLM Inference,'' \textit{IEEE Micro}, vol.~44, no.~3, 2024. doi: 10.1109/MM.2024.3375352.

\bibitem{cxlspeckvl2024}
D.~Liu and Y.~Yu, ``CXL-SpecKV: A Disaggregated FPGA Speculative KV-Cache for Datacenter LLM Serving,'' arXiv:2512.11920, 2025.

\bibitem{nvidia_h100}
NVIDIA Corporation, ``NVIDIA H100 Tensor Core GPU Architecture,'' Tech.\ Rep., 2023.

\bibitem{gpudirectstorage}
NVIDIA Corporation, ``GPUDirect Storage: A Direct Path Between Storage and GPU Memory,'' Tech.\ Rep., 2023.

\bibitem{kalia2014rdma}
A.~Kalia, M.~Kaminsky, and D.~G.~Andersen, ``Using RDMA Efficiently for Key-Value Services,'' in \textit{Proc.\ ACM SIGCOMM}, pp.~295--306, 2014.

\bibitem{murphy2012mlpp}
K.~P.~Murphy, \textit{Machine Learning: A Probabilistic Perspective}. MIT Press, 2012.

\bibitem{thompson1933sampling}
W.~R.~Thompson, ``On the Likelihood that One Unknown Probability Exceeds Another in View of the Evidence of Two Samples,'' \textit{Biometrika}, vol.~25, no.~3/4, pp.~285--294, 1933.

\bibitem{guo2016rdma}
C.~Guo \textit{et~al.}, ``RDMA over Commodity Ethernet at Scale,'' in \textit{Proc.\ ACM SIGCOMM}, pp.~202--215, 2016.

\bibitem{patel2024splitwise}
P.~Patel \textit{et~al.}, ``Splitwise: Efficient Generative LLM Inference Using Phase Splitting,'' in \textit{Proc.\ ISCA}, pp.~118--132, 2024.

\bibitem{zhong2024distserve}
Y.~Zhong \textit{et~al.}, ``DistServe: Disaggregating Prefill and Decoding for Goodput-optimized Large Language Model Serving,'' in \textit{Proc.\ OSDI}, pp.~193--210, 2024.

\bibitem{xu2024kvcachesurvey}
H.~Li \textit{et~al.}, ``A Survey on Large Language Model Acceleration based on KV Cache Management,'' arXiv:2412.19442, 2024.

\bibitem{nvidia_dynamo2024}
NVIDIA Corporation, ``NVIDIA Dynamo: Dynamic GPU Inference Serving,'' Tech.\ Rep., 2025.

\bibitem{sharegpt2023}
ShareGPT Community, ``ShareGPT: Sharing ChatGPT Conversations,'' 2023. \url{https://sharegpt.com}.

\bibitem{zheng2024lmsyschat}
L.~Zheng \textit{et~al.}, ``LMSYS-Chat-1M: A Large-Scale Real-World LLM Conversation Dataset,'' arXiv:2309.11998, 2024.

\bibitem{ye2026pkas}
J.~Ye, A.~Maurya, K.~T.~Chitty-Venkata, B.~Nicolae, A.~Kougkas, and X.-H.~Sun, ``PKAS: Predictive KVCache-Aware Scheduling for Faster LLM and Transformer Inferences,'' in \textit{Proc.\ HPDC}, 2026. doi: 10.1145/3806645.3807819.

\bibitem{ao2026congestion}
R.~Ao, J.~Dong, G.~Luo, and D.~Simchi-Levi, ``Service-Induced Congestion in Memory-Constrained LLM Serving,'' arXiv:2606.15555, 2026.

\bibitem{yu2026crosstier}
D.~Yu, ``Cost-Efficient Multimodal LLM Inference via Cross-Tier GPU Heterogeneity,'' arXiv:2603.12707, 2026.

\end{thebibliography}
\end{document}